# Precision quantization of Hall resistance in transferred graphene

M. Woszczyna, M. Friedemann, M. Götz, E. Pesel, K. Pierz, T. Weimann, F. J. Ahlers*

Physikalisch-Technische Bundesanstalt, Bundesallee 100, D-38116 Braunschweig, Germany

*Correspondence to: franz.ahlers@ptb.de

We show that quantum resistance standards made of transferred graphene reach the uncertainty of semiconductor devices, the current reference system in metrology. A large graphene device (150 × 30 μm$^2$), exfoliated and transferred onto GaAs, revealed a quantization with a precision of $(-5.1 \pm 6.3) \cdot 10^{-9}$ accompanied by a vanishing longitudinal resistance at current levels exceeding 10 μA. While such performance had previously only been achieved with epitaxially grown graphene, our experiments demonstrate that transfer steps, inevitable for exfoliated graphene or graphene grown by chemical vapor deposition (CVD), are compatible with the requirements of high quality quantum resistance standards.

Our physical world is described by measurements. To reproduce the units of measurements independent of measurement condition, as precisely as possible and with methods available to everyone, is the ultimate quest of metrology. The perfect route proceeds via the fundamental constants of nature by exploiting quantum mechanical effects. A most prominent example is the electrical resistance which can be reproduced by the ratio of Planck's constant $h$ and the elementary charge $e$ as $h/e^2 \approx 25.8$ kΩ. This exact resistance quantization is known as the quantum Hall effect (QHE) and is observed in two-dimensional electron systems (2DES), such as semiconductor heterostructures grown by molecular beam epitaxy[1]. Graphene[2] has become an attractive new candidate because it is two-dimensional by nature, can be prepared by simple peeling from natural graphite and subsequent transfer onto an insulating substrate, and exhibits a huge cyclotron energy splitting in a magnetic field, which makes the QHE observable even at room temperature[3]. This latter fact was the key trigger for several national metrology institutes to start intensive work on realizing a quantum resistance standard working at such high temperature and low magnetic field that its dissemination becomes as simple as that of Josephson voltage standards which have already found their way into industrial laboratories many years ago[4,5].

By epitaxial techniques one may obtain graphene devices revealing a quantization uncertainty as low as a few parts in $10^9$ [6], matching traditional standards. Furthermore, graphene has passed a universality test[7], in which a direct comparison between quantized Hall resistances in graphene and semiconductor devices gave exactly the same value within an uncertainty of only 9 parts in $10^{11}$. The key for this achievement was a large device area (160 μm by 35 μm) in which quantum transport channels in the 2DES were sufficiently separated in space to maintain high electric currents, allowing the high signal-to-noise ratio required for low uncertainty measurements. Although the epitaxial technology provides even wafer scale graphene sheets, it is challenging, requires sophisticated facilities, and the graphene cannot be readily combined with other substrates. A further complication arises from the difficulty of reducing and tuning its very high carrier concentration, which is an important prerequisite for an application as a resistance standard.

On the other hand, graphene has taken science by storm mainly due to its ease of fabrication. "Kitchen table"-made tuneable graphene devices are available[8], in which carrier densities are set by simply applying bias voltage to the back side of an insulating substrate. A drawback is that graphene flakes with dimensions of at most tens of micrometers are typically obtained, perfect for research but too small for metrology application. Indeed, early attempts to prove the usefulness of the simply made graphene for metrology only showed a relative measurement uncertainty of 15 parts in $10^6$ [9], orders of magnitude worse than what is possible with semiconductors. Even the current state-of-the-art precision measurement of QHE in exfoliated graphene revealed, after careful data evaluation, only a measurement uncertainty of 5 parts in $10^7$ [10].

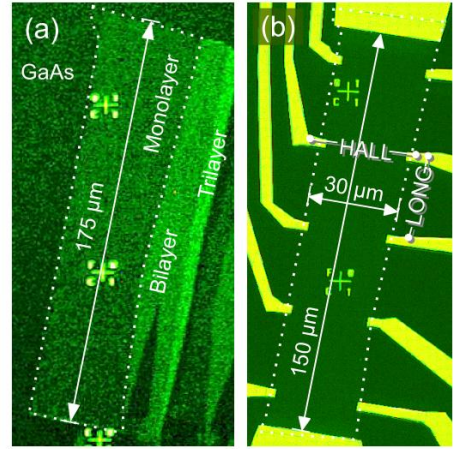

FIG. 1. (Color online) Large area unprocessed graphene flakes on GaAs substrate (a) and a complete graphene device (b) made from the monolayer marked by the dotted line.

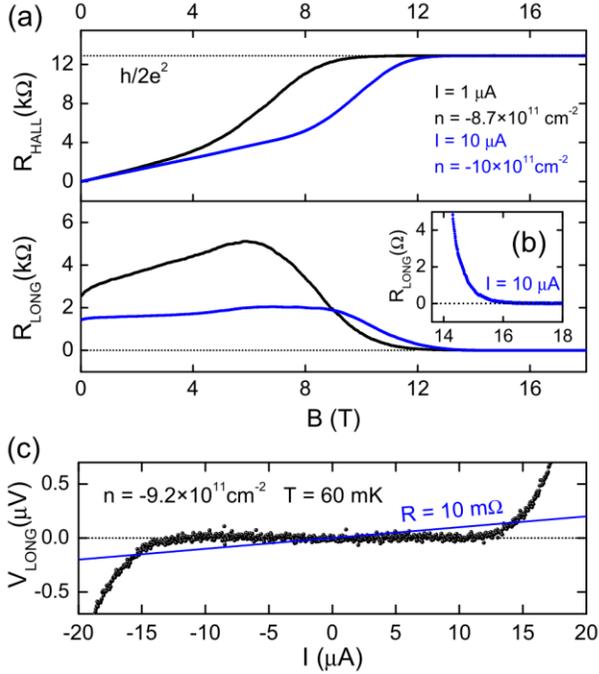

FIG. 2. (Color online) (a) High current quantum Hall effect in graphene on GaAs showing Hall (upper panel) and longitudinal (lower panel) resistances in dependence on the magnetic field. Data for two different carrier concentrations were taken in different cool-down cycles. The concentrations were determined from low field Hall measurements. Inset (b) shows a magnified longitudinal resistance plot for a magnetic field range from 13.8 T to 18 T. (c) Longitudinal voltage drop in dependence on the supply current at the magnetic field 18 T. The sloped line represents a threshold resistance of 10 mΩ.

In this communication, we show that one can obtain large graphene devices suitable for precision metrology at reasonable yield and that the technology steps required for transfer do not compromise this suitability. In order to achieve this, we chose a substrate other than the commonly employed Si/SiO$_2$ material used nearly exclusively for exfoliated graphene. Our choice of GaAs was initially triggered by two expectations. Firstly, the surface roughness of GaAs is lower than that of thermally grown SiO$_2$ and should thus favor a higher quality of the graphene, and secondly, its higher dielectric constant should improve electrical screening of substrate defects[11]. It later turned out that as an additional advantage, a higher yield of rather large graphene devices enabled the decisive breakthrough, namely the fabrication of Hall bars with dimensions matching those made from SiC-grown graphene. We speculate that it is the stronger hydrophilic character of GaAs which leads to a better "stickiness" of graphene flakes and prevents their folding over, which occurs with flakes on SiO$_2$, and limits the obtainable flake size. After a special technique had been developed which made graphene flakes on GaAs visible[12] and at the same time provided a back-gate insulator[11], electrically tuneable Hall bar devices could be fabricated. Fig. 1(a) depicts graphene flakes transferred on GaAs, visible even to the naked eye, and Fig. 1(b) shows the final device fabricated by conventional technology steps comprising lithography and Ti-Au contact metal evaporation. At a moderate carrier mobility of 3000 cm$^2$/Vs the sample was not perfectly homogeneous, and we restricted our investigations to the contact pairs depicted in the inset in Fig. 1(a). Note that two of the metal markers deposited on the surface after exfoliation for fabrication alignment lay on the graphene flake. However, a wide and perfectly developed quantum Hall resistance plateau at filling factor ν = -2 (Fig. 2(a), upper panel) accompanied by vanishing longitudinal resistance (Fig. 2(b), lower panel) was observed at a temperature of 60 mK even at high currents of several microamperes. Plateaus at higher filling factors were barely visible. In order to limit back-gate leakage currents, we further restricted the investigation to the ν = -2 plateau, for which the back-gate voltage, defining the graphene carrier concentration, was around 0 V. Fig. 2(c) presents the longitudinal voltage drop along the Hall bar in dependence on supply current at a magnetic field of 18 T. Contact resistances at the plateau were below 10 Ω, and the onset of the QHE breakdown occurred only for currents exceeding ±10 µA. (Note that a recent study identified high energy loss rates of hot carriers in exfoliated graphene to be responsible for the high breakdown current density[13]).

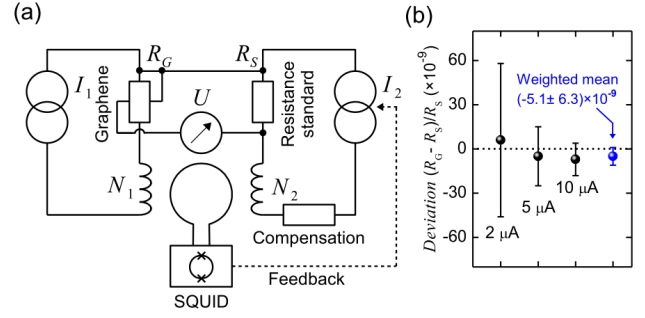

FIG. 3. (Color online) (a) Simplified scheme of the precision QHE measurement utilizing a CCC setup. (b) Relative deviations between quantum Hall resistance in graphene and a conventional resistance standard for different currents (in black). The final measurement result (in blue) revealed, within the measurement uncertainty of a few parts in $10^9$, zero resistance deviation.

To quantitatively compare the graphene performance with a semiconductor quantum resistance standard, a cryogenic current comparator (CCC) bridge was used[14], the most sensitive apparatus for high-precision resistance comparisons. The CCC allowed resolving resistance differences with a measurement uncertainty below 1 part in $10^{10}$. We compared the graphene Hall resistance $R_G$ with the resistance of a precisely known standard 100-Ω resistor $R_S$, which was calibrated against the semiconductor QHE standard just before and after the graphene device measurements. A simplified scheme of the experiment is presented in Fig. 3(a). The resistance ratio $R_G/R_S$ was determined from the ratio of the DC currents $I_1$ and $I_2$ and from the imbalance of the bridge $U$. The ultrasensitive SQUID electronic unit controlled $I_2$ in order to keep the ratio $I_2/I_1$ equal to the exactly known winding ratio $N_1/N_2$. An auxiliary compensation network was used for obtaining non-integer current ratios[10]. Typical results obtained at current levels $I_1$ of 2, 5 and 10 µA are shown in Fig. 3(b), where *Deviation* is calculated as a relative difference between the quantized Hall resistance measured in graphene and the value of $h/2e^2$ provided by the semiconductor device. Error bars indicate $1\sigma$ statistical standard deviation and represent the total measurement uncertainty. The behavior of the graphene Hall bar was the same as that of a conventional GaAs/AlGaAs heterostructure device: increasing measurement current led to decreasing measurement uncertainty for the constant CCC integration time (22 minutes in this case), as long as the current level was below the QHE-breakdown threshold. The weighted average of the collected *Deviation* data was found to be compatible with zero within the uncertainty of 6.3 parts in $10^9$, meaning that this graphene sample fulfills the stringent requirements for electrical resistance standards, previously met only by semiconductor and epitaxial graphene devices.

To determine the QHE-breakdown current threshold we use the same definition as in [15] where it had been defined as that current where the longitudinal voltage deviates from zero within a measurement noise of 10 nV. For our sample this was the case for a current of 12 µA. However, to easier compare our breakdown behavior with longitudinal voltage values reported in [6,7,13], we show in Fig. 2(c) a line representing a resistance of 10 mΩ. From this line we obtain a breakdown current density of 15µA/30µm = 0.5 A/m, the same value as in [6] where it can be deduced from Fig. 2(d). In [7] a much higher value of 14 A/m has been reported for polymer gated epitaxial graphene, and recently 1.3 A/m were reported for exfoliated graphene in the non-annealed state of a 20 µm wide Hall bar [13], although in that work no high precision Hall resistance measurements were made.

In this work, we have proven that transferred graphene can compete with epitaxially grown graphene in quantum metrology applications. In our case, an exceptionally large exfoliated graphene device was fabricated on GaAs. The significance of the result is in demonstrating that in a precision metrology application transferred graphene can be used, provided the fabrication process allows obtaining Hall bar sizes exceeding dimensions of tens of micrometers. Many groups have already mastered techniques of transferring graphene onto various substrates, the most promising of which is boron nitride[16]. The remaining yield constraints of transferred graphene will likely be relieved by the progress of CVD-growth[17] of graphene. All this suggests that one may soon expect primary quantum resistance standards to become accessible to a wider community in science and industry.